\documentclass{emulateapj}
\def\wfig{1.0}\def\wfigb{0.95}             

\usepackage{amsbsy,natbib}

\usepackage{amssymb}
\usepackage{amsmath}
\usepackage{amsthm}
\usepackage{amsfonts}

\graphicspath{{fig/}}

\newcommand{\f}[2]{#1$\,$10$^{#2}$}        

\newcommand{\Fig}[1]{Fig.\ \ref{fig:#1}}
\newcommand{\Figure}[1]{Figure \ref{fig:#1}}

\begin{document}

\title{Abundance of $^{\bf26}$Al and $^{\bf60}$Fe in evolving Giant Molecular Clouds}

\author{Aristodimos Vasileiadis\altaffilmark{1,2}, \AA ke Nordlund\altaffilmark{1,2} and Martin
Bizzarro\altaffilmark{1,}}

\altaffiltext{1}{Centre for Star and Planet Formation, Natural History Museum of
Denmark, University of Copenhagen, \O stervoldgade 5-7, 1350 Copenhagen \O,
Denmark }
\altaffiltext{2}{Niels Bohr Institute, University of Copenhagen, Blegdamsvej 17,
2100 Copenhagen, Denmark}

\begin{abstract}  
The nucleosynthesis and ejection of radioactive $^{26}$Al (t$_{1/2} \sim$ 0.72\,Myr) and $^{60}$Fe, (t$_{1/2} \sim$ 2.5\,Myr) into the interstellar medium is dominated by the stellar winds of massive stars and supernova type II explosions. Studies of meteorites and their components indicate that the initial abundances of these short-lived radionuclides in the solar protoplanetary disk were higher than the background levels of the galaxy inferred from $\gamma$--ray astronomy and models of the galactic chemical evolution. This observation has been used to argue for a late-stage addition of stellar debris to the Solar System's parental molecular cloud or, alternatively, the solar protoplanetary disk, thereby requiring a special scenario for the formation of our Solar System. Here, we use supercomputers to model---from first principles---the production, transport and admixing of freshly synthesized $^{26}$Al and $^{60}$Fe in star-forming regions within giant molecular clouds. Under typical star-formation conditions, the levels of $^{26}$Al in most star-forming regions are comparable to that deduced from meteorites, suggesting that the presence of short-lived radionuclides in the early Solar System is a generic feature of the chemical evolution of giant molecular clouds. The $^{60}$Fe/$^{26}$Al yield ratio of $\approx 0.2$ calculated from our simulations is consistent with the galactic value of 0.15 $\pm$ 0.06 inferred from $\gamma$--ray astronomy but is significantly higher than most current Solar System measurements indicate. We suggest that estimates based on differentiated meteorites and some chondritic components may not be representative of the initial $^{60}$Fe abundance of the bulk Solar System.

\end{abstract}

\keywords{ ISM: kinematics and dynamics --- stars: formation --- nuclear reactions, nucleosynthesis, abundances --- protoplanetary disks --- meteorites, meteors, meteoroids}

\section{Introduction}

Due to the low opacity of the Milky Way to $\gamma$-ray emissions from the radioactive $^{26}$Al and $^{60}$Fe nuclei, $\gamma$-ray observations provide estimates of the mean $^{60}$Fe and $^{26}$Al abundances in the Galaxy. Combined with galactic chemical evolution models, these observations can be used to determine the steady-state abundance of $^{26}$Al and $^{60}$Fe in the interstellar medium and, importantly, a robust estimate of the $^{60}$Fe/$^{26}$Al ratio of the Galaxy \citep{2006Natur.439...45D,2007A&A...469.1005W}. Meteorites and their components, including the Solar System's first solids---calcium-aluminum-rich inclusions (CAIs)---contain evidence for an early presence of $^{26}$Al and $^{60}$Fe at levels that were apparently higher than expected from the galactic background abundance, especially for $^{26}$Al \citep{2009GeCoA..73.4963K,2009GeCoA..73.4922H}. Moreover, the variable initial abundances of $^{10}$Be, a short-lived radionuclide uniquely synthesized by spallation reactions, in CAIs recording uniform initial $^{26}$Al/$^{27}$Al values indicate that no more than $\sim$2.5\% of the solar system's $^{26}$Al inventory may have been produced by local irradiation processes \citep{2012ARA&A..50..531K}. These observations are interpreted as reflecting a late-stage contamination of the nascent Solar System from a nearby supernova \citep{2004NewAR..48..125G}. However, the majority of Sun-like stars form in cold and dense molecular cloud regions encapsulated in Giant Molecular Cloud (GMC) structures \citep{2012ARA&A..50..531K}, an astrophysical environment different from that of the average galactic interstellar medium (ISM). Therefore, an overall galactic abundance level of $^{26}$Al and $^{60}$Fe is not necessarily representative of levels encountered in star-forming regions, where the cold and dense gas out of which stars form may inherit material enriched in freshly synthesized matter by nearby, recent generations of stars.

To better understand the abundance levels of $^{26}$Al and $^{60}$Fe in star-forming regions, we used three-dimensional magnetohydrodynamic models of self-gravitating GMCs to simulate the time-integrated production and ejection of these radioactive species via the supernova and stellar wind mechanisms of ensembles of stars, tracking their incorporation into star-forming clumps. In these simulations, the ISM is assumed to be driven by the kinetic energy originating from supernova input, as well as from larger, galactic-scale, density perturbations, producing giant molecular clouds with morphology and kinematics controlled by gravo-turbulent fragmentation. A hierarchical structure results, with GMCs containing smaller scale cold Molecular Cloud (MC) fragments, which in turn contain filamentary structures where sufficiently dense and massive pre-stellar cores may become gravitationally unstable and collapse into new-born stars, with initial masses distributed according to an Initial Mass Function \citep[IMF,][]{2002ApJ...576..870P,2005ASSL..327...41C}. Sufficiently massive stars explode as supernovae after a few million years, enriching their surroundings with freshly synthesized metals as well as contributing kinetic energy back into the ISM.

\section{Methods}
We use periodic simulation boxes with sizes ranging from 40--80 pc, filled with self-gravitating, magnetized gas of average mass density consistent with Larson's mass-size relation \citep{1981MNRAS.194..809L}; this results in $H_2$ number densities between 30 and 15 $\,$cm$^{-3}$, corresponding to total masses $\sim 10^{5}$ and \f{4\,$\cdot$}{5} $M_{\odot}$ respectively. Experiments are initialized with the unigrid magneto-hydrodynamics {{\sc stagger}} code \citep[see][]{2011ApJ...737...13K} that includes statistical star formation and individual supernova explosions, with tracking of yields \citep[][]{2006ApJ...647..483L} using a passive scalar transport equation. The passive scalar variable is assigned the decay time of its representing isotope \citep[$^{26}$Al decays to $^{26}$Mg with a half-life of 0.72 Myr and $^{60}$Fe decays to $^{60}$Ni with a half-life of 2.62 Myr;][]{1983JGR....88..331N,2009PhRvL.103g2502R}. Solar abundances \citep[][]{2009ARA&A..47..481A} are assumed for the stable reference isotopes $^{56}$Fe and $^{27}$Al. Bit-wise identical simulation runs are executed in pairs, differing only in half-life and yields. Because the passive scalar does not affect the dynamic evolution of the simulation short-lived radioisotope (SLR) data from different runs can be superimposed.

The periodic boundary conditions represent a GMC fragment surrounded by similar or larger amounts of nearby gas, with the surrounding gas pressure preventing break-up due to supernova explosions and other feedback processes; GMC life times are consistent with the `star formation in a crossing time' paradigm \citep{2000ApJ...530..277E,2013IAUS..292...35E}. The view that feedback is very important for cloud breakup, while popular, is actually not supported by convincing, large scale numerical experiments \citep{2011arXiv1111.2827K}. Inertial forcing from larger scales is replaced our model by solenoidal forcing in a shell of low wavenumbers \citep[][]{2002ApJ...576..870P}, here $1/L \leq k/2\pi \leq 2/L$, producing velocity dispersions of the cold and dense medium  consistent with Larson's velocity relation \citep{1981MNRAS.194..809L}. Heating and cooling is modeled as schematic heating by UV-photons \citep{1989agna.book.....O} quenched in dense gas according to the recipe of \citet{1986PASP...98.1076F}, and an optically thin cooling function consistent with \citet{2012ApJS..202...13G}. Connected cold and dense regions of sufficient mass ($>10^3\,M_{\odot}$ for occasional formation of $10^2\,M_{\odot}$ stars) are converted to stars at a given star formation rate, $\varepsilon$, with an IMF-distribution of masses consistent with \citet{2002ApJ...576..870P}. The star formation rate per unit free fall time $\varepsilon$ was varied from $0.02$ to $0.1$, covering a range consistent with observations, numerical models, and theory \citep{2011ApJ...729..133M,2011ApJ...730...40P,2005ApJ...630..250K}.

Stars $>$ 8 $M_{\odot}$ are followed until they explode as Type II supernovae, after lifetimes interpolated from \citet{1994A&AS..103...97M}.  Supernovae are assumed to release $10^{51}$ erg of thermal energy, distributed in a mass corresponding to the entire mass of the progenitor star. We combine the small expected integrated wind yields into the generally larger supernova yields. Supernovae explosions result in initial temperatures of $\sim$$10^8$ K and initial expansion velocities of $\sim$1000--3000 km s$^{-1}$, which for brief periods of time enforce time steps $\sim$years. To allow the  initial state of constant density and zero velocity to develop into realistic GMC initial conditions, we let the forcing generate a supersonic velocity field with average velocity dispersion consistent with \citet{1981MNRAS.194..809L}, and wait until density fluctuations generate a first generation of stars, where the most massive stars already begin to explode as supernovae.

When supernova driving of the medium becomes significant a snapshot is passed to a locally-modified version of the adaptive mesh refinement code {\sc ramses} \citep{2002A&A...385..337T,2005sf2a.conf..743F}, with stellar lifetimes, supernova yields, and ISM heating and cooling consistent with that used in the {\sc stagger} parts of the experiments. In {\sc ramses} experiments there is no external driving and star formation is handled individually, with new-born stars represented by accreting sink particles \citep{2011ApJ...730...40P,2012ApJ...759L..27P}. The {\sc ramses} code allows for tracking the transport of several nuclides in a single simulation. Massive stars are tracked until they explode as Type II supernovae. We use a $128^3$ root grid and refine to a local resolution corresponding to $4096^3$ (cell size at the highest refinement level $\Delta s \approx 0.01$ pc $\approx 2000$ AU). Tests with $256^3$ {\sc ramses} root grids and corresponding {\sc stagger} code resolutions did not give significantly different results.

\section{Enhanced abundances of $^{\bf26}$Al and $^{\bf60}$Fe in evolving GMCs}

\begin{figure}[htb!]
  \centerline{\includegraphics[width=\wfig\linewidth]{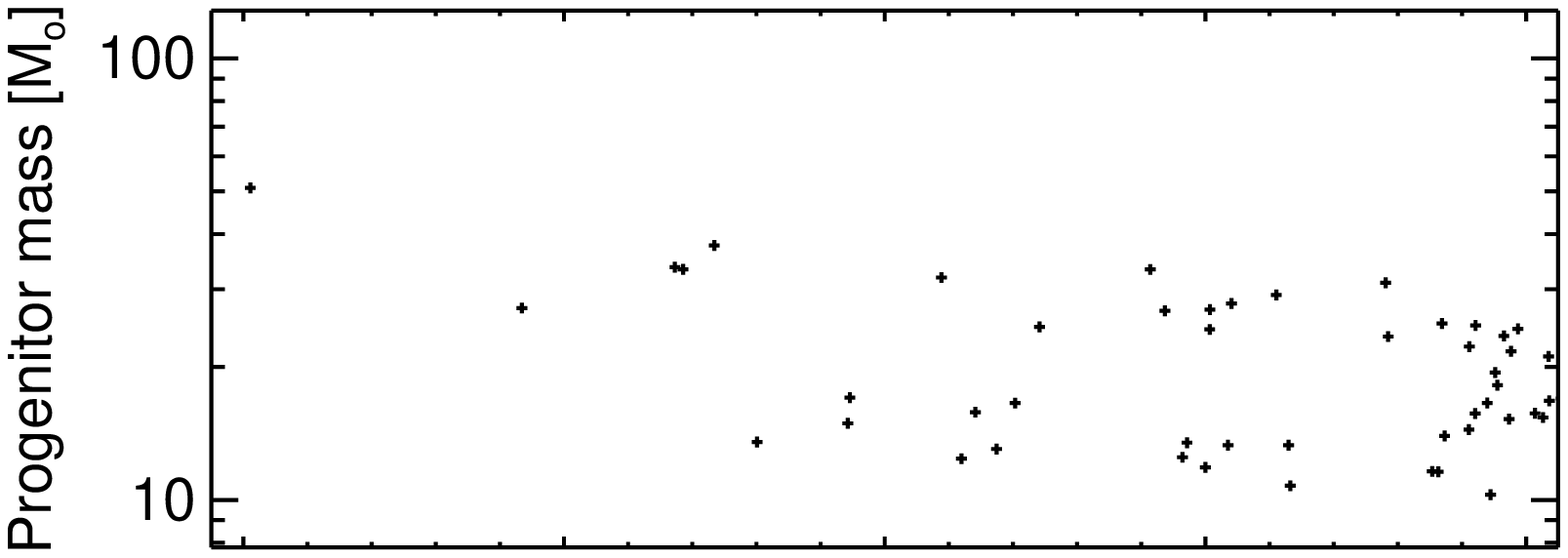}}
  \vspace*{-1cm}
  \centerline{\includegraphics[width=\wfig\linewidth]{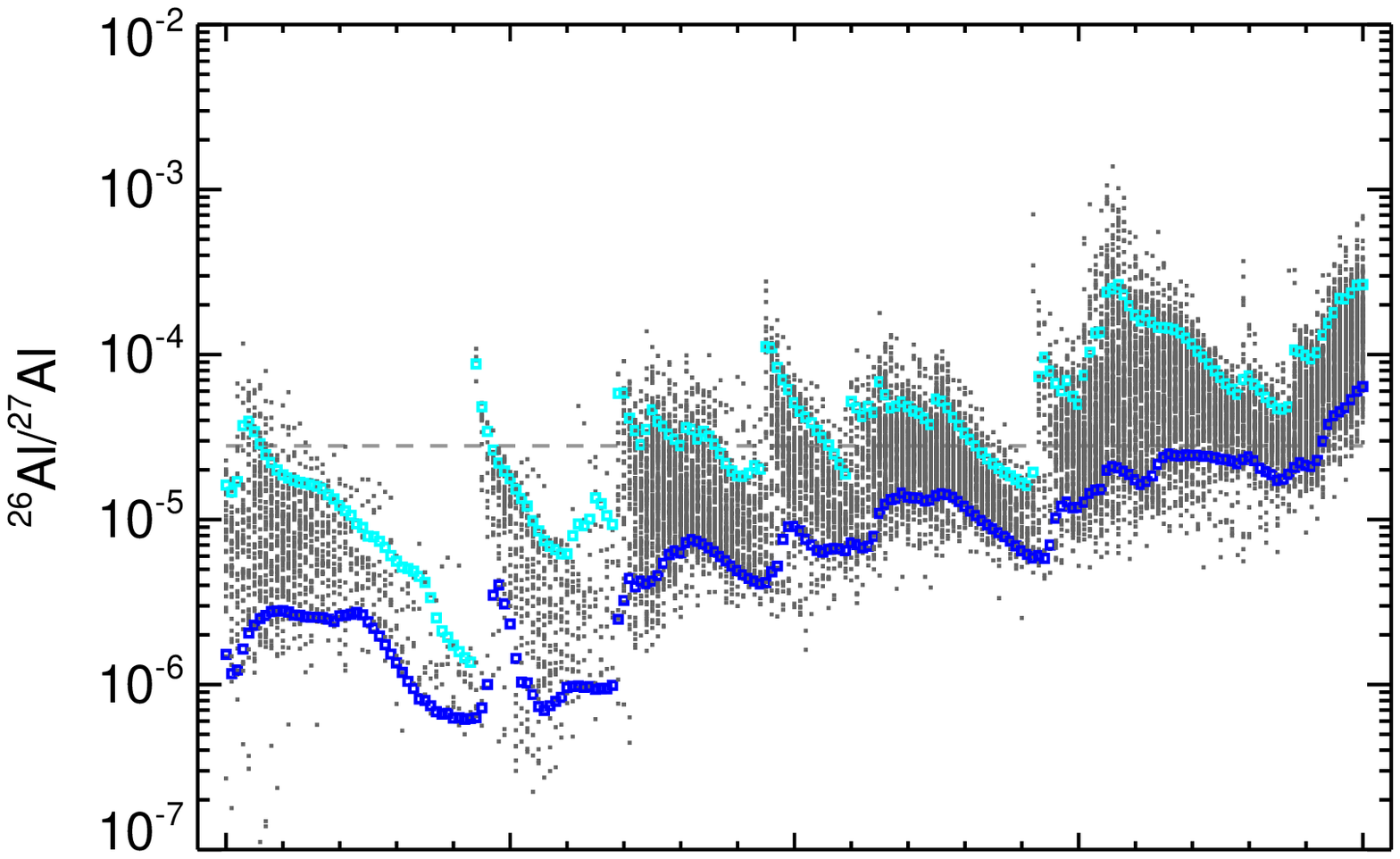}}
  \vspace*{-1cm}
  \centerline{\includegraphics[width=\wfig\linewidth]{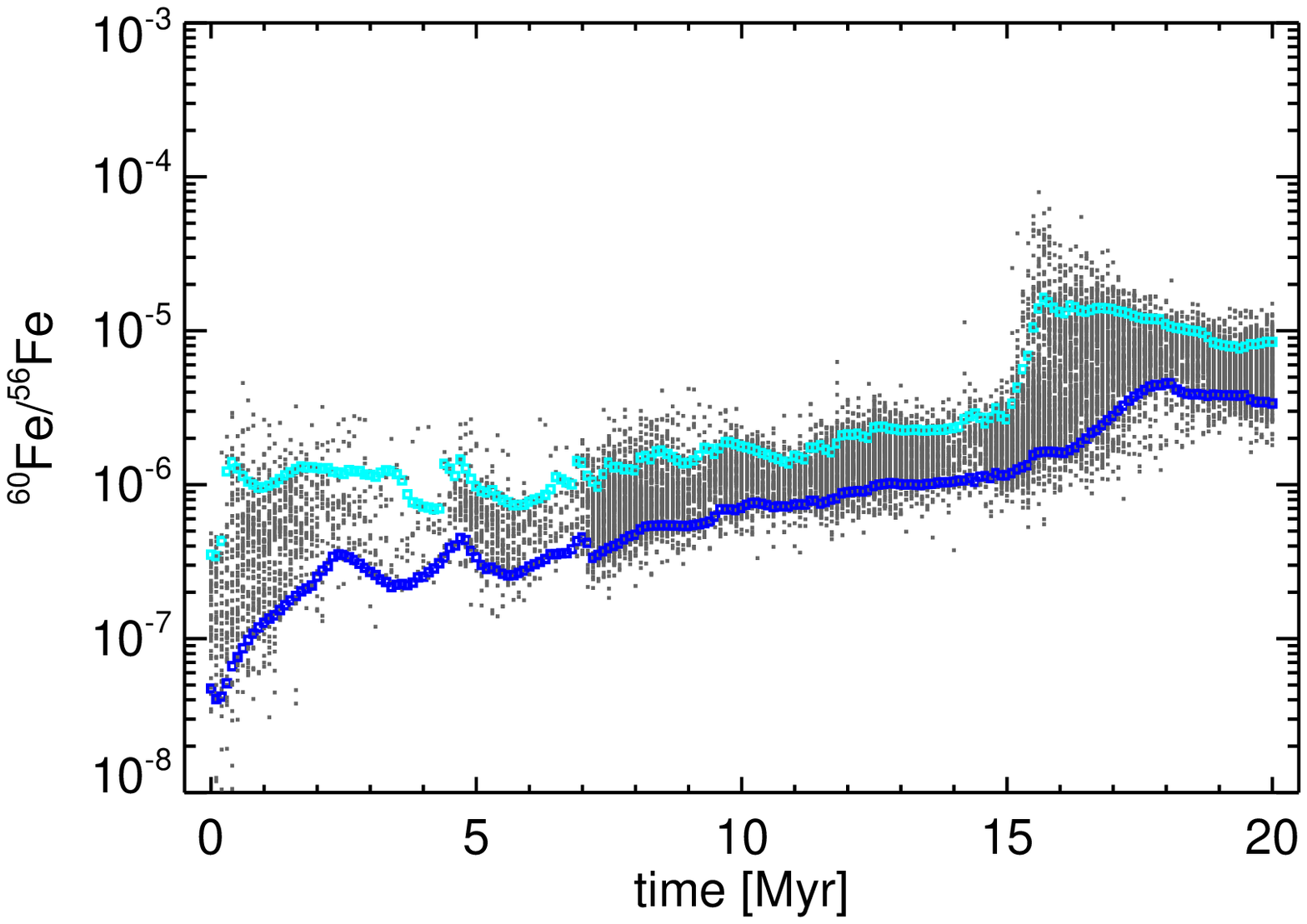}}
  \caption{Supernova mass as a function of time (top), and values of the ratios
  $^{26}$Al/$^{27}$Al (mid) and $^{60}$Fe/$^{56}$Fe (bottom) in dense, star-forming gas
  of an evolving 40 pc, 10$^{5}$ M$_{\odot}$ GMC with $\varepsilon=0.05$. Grey dots show values sampled in dense,
  star-forming clumps, with 10\% and 90\% percentiles marked in blue and cyan. Dashed line is the modeled bulk solar value of 
  $2.8\cdot10^{-5}$ \citep{2011ApJ...735L..37L}. Time is
  indicated in Myr after formation of the GMC (10 Myr offset from simulation start).}
  \label{fig:1}
\end{figure}

We find that $\sim$10 Myr of evolution is needed for the model GMC to develop properties similar to young GMCs, where most of the massive stars are waiting to explode as supernovae. Over the next $\sim$20 Myr, the results show a gradual increase of supernova feedback, with levels of both kinetic energy and average yields behaving in a similar way. This time interval represents the lifetime of the model GMC where we track the production, transport and admixing of stellar-derived $^{26}$Al and $^{60}$Fe into cold and dense star-forming gas. Over this time period, consistent with estimates of GMC lifetimes \citep{2011ApJ...729..133M}, the average concentration of $^{26}$Al and $^{60}$Fe in star-forming gas increases, with $^{26}$Al/$^{27}$Al and $^{60}$Fe/$^{56}$Fe values ranging from $\sim$10$^{-6}$ to $\sim$10$^{-4}$ and $\sim$10$^{-7}$ to $\sim$10$^{-5}$, respectively (Fig.\ \ref{fig:1}). There is significant variability in the $^{26}$Al/$^{27}$Al and $^{60}$Fe/$^{56}$Fe ratios amongst the individual star-forming clumps interspersed within the GMC throughout its evolution. A first important quantitative result is that the concentration of $^{26}$Al and $^{60}$Fe in dense star-forming gas is systematically higher than the inferred galactic background abundances of these radioactive species, estimated to  $^{26}$Al/$^{27}$Al and $^{60}$Fe/$^{56}$Fe values of $\sim2.3\cdot10^{-6}$ and 
$\sim1.6\cdot10^{-8}$, respectively \citep{2009GeCoA..73.4922H}. Transient bursts of $^{26}$Al and $^{60}$Fe abundances observed in the first 10 Myr of the simulation predominantly reflect single supernova events of the most massive stars ($M_{\odot}$$>$20). In later stages of the simulation ($>$10 Myr), most of the $^{26}$Al and $^{60}$Fe budget is contributed from the longer-lived lower mass massive stars (8$<$$M_{\odot}$$<$20), typified by more frequent lower yields compared to their rarer and more massive counterparts. This results in a more flat distribution of $^{26}$Al and $^{60}$Fe amongst star-forming regions.

\begin{figure}[htb!]
  \centerline{\includegraphics[width=\wfig\linewidth]{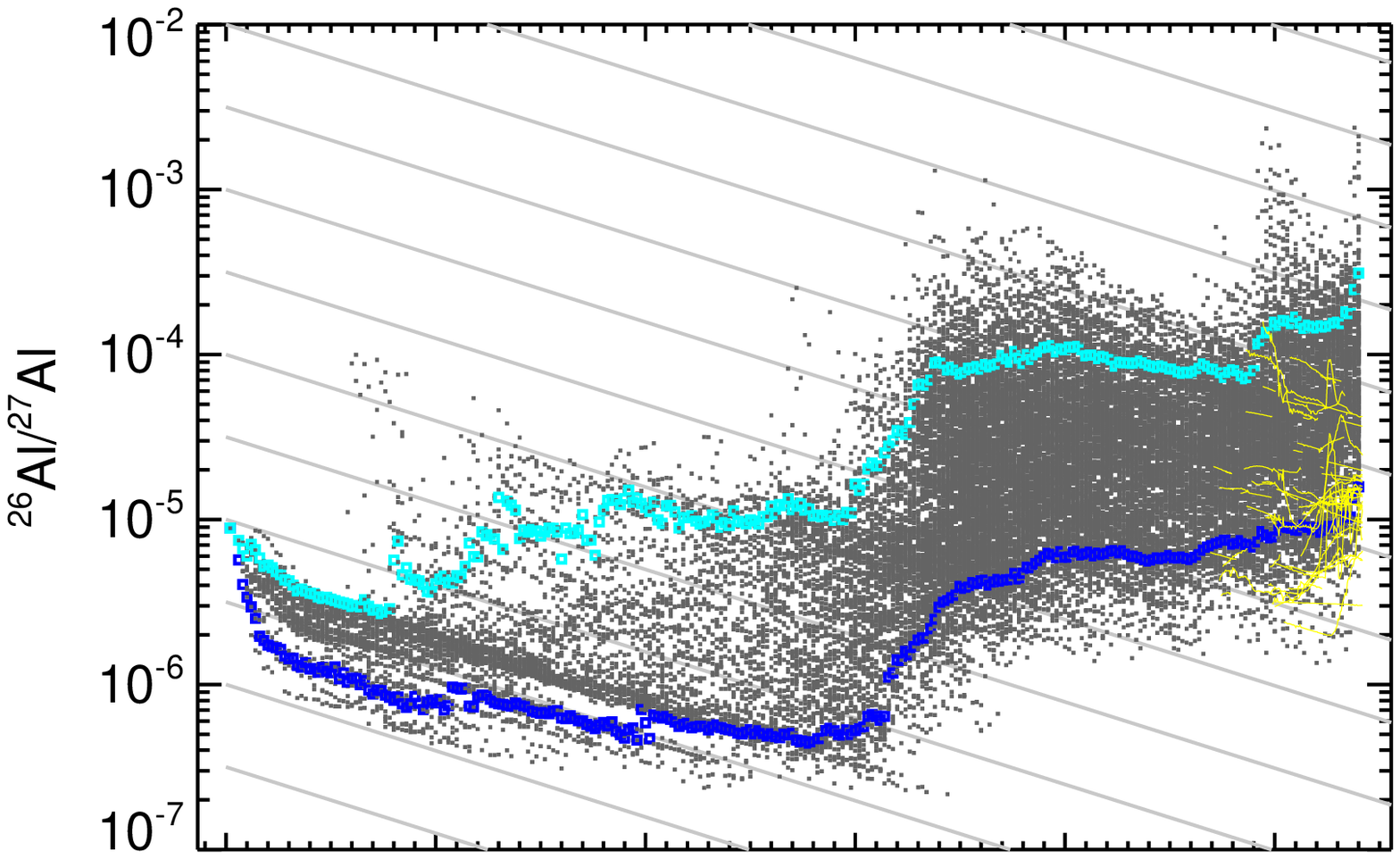}}
  \vspace*{-1cm}
  \centerline{\includegraphics[width=\wfig\linewidth]{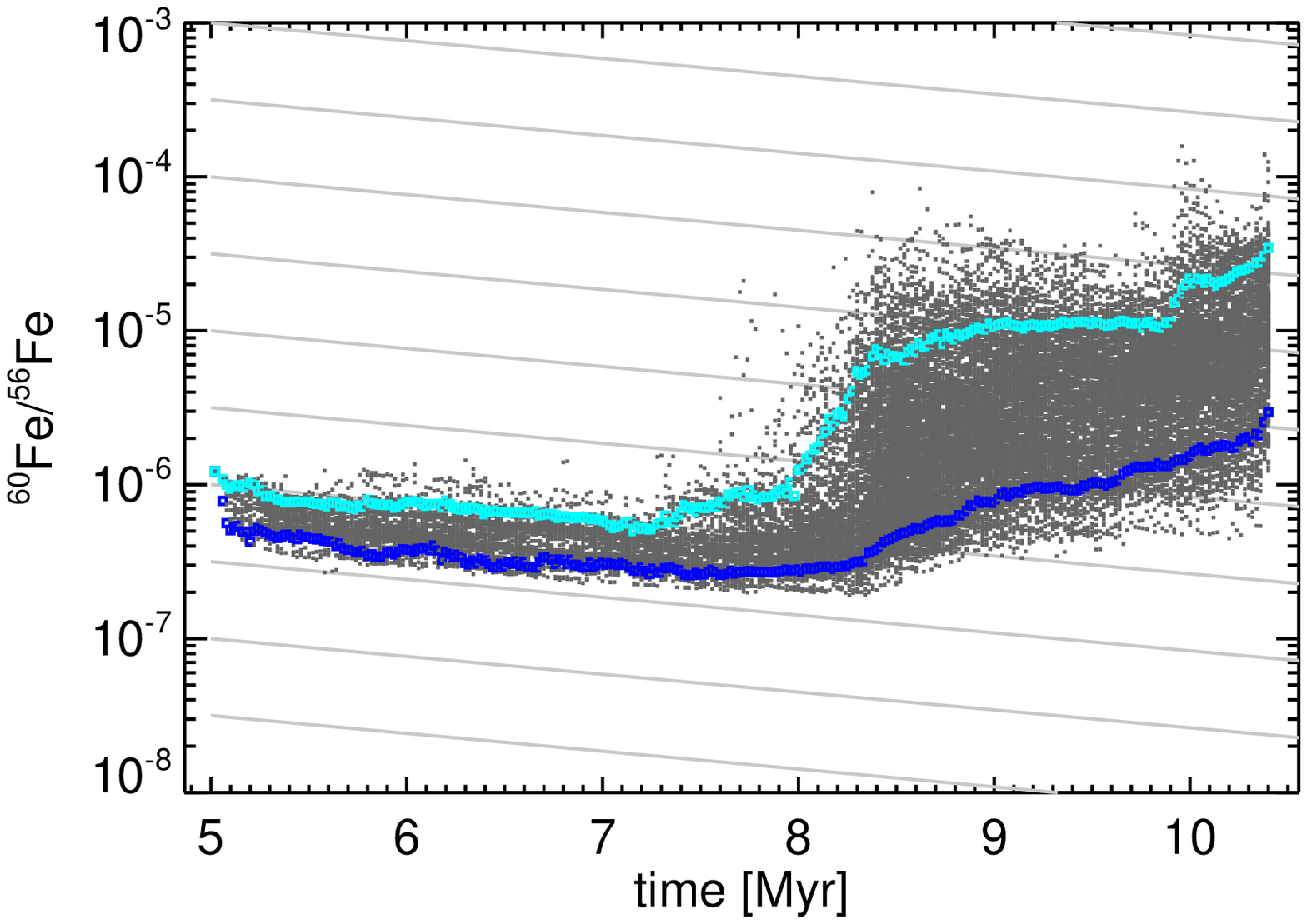}}
  \caption{Similar results as those shown in the middle and bottom panels of Fig.\ \ref{fig:1},
  but computed with the {{\sc ramses}} adaptive mesh refinement code. The timescale corresponds to that
  of Fig.\ \ref{fig:1}. The slanted gray lines in background indicate the decay
  rates. Yellow thin curves show instantaneous relative $^{26}$Al/$^{27}$Al abundance in gas accreting to sink particles representing individual stars.}
  \label{fig:2}
\end{figure}

\Figure{2} shows {\sc ramses} results for the 5--10 Myr interval of the model GMC, computed with a finest grid resolution of 0.01 pc, which allows us to trace the concentration of $^{26}$Al in cold and dense gas accreting to individual stars. The average $^{26}$Al abundance during that time period is comparable to that obtained with the low-resolution  {\sc stagger} code although the variability of the $^{26}$Al/$^{27}$Al values is greater in the high-resolution {\sc ramses} case, which is able to maintain larger amplitude inhomogeneities inside the compact molecular cloud structures. We note that gas accreting onto our 'sink particle' representation of stars do so at densities $\sim 10^{7}$ cm$^{-3}$. These contributions shows similar levels of $^{26}$Al---within the rather large statistical spread--as that of the average star forming gas, sampled at $\sim 10^{4}$ cm$^{-3}$. This illustrates that, although inefficient, star-formation in cold molecular clouds is a locally rapid process, occurring on time scales of order crossing times \citep{2000ApJ...530..277E}. Thus, we infer that the enhanced abundance of $^{26}$Al and $^{60}$Fe in star-forming gas observed in our simulations---as compared to the average galactic abundances---is a generic feature of the chemical evolution of GMCs.

\begin{figure}[htb!]
  \centerline{\includegraphics[width=\wfig\linewidth]{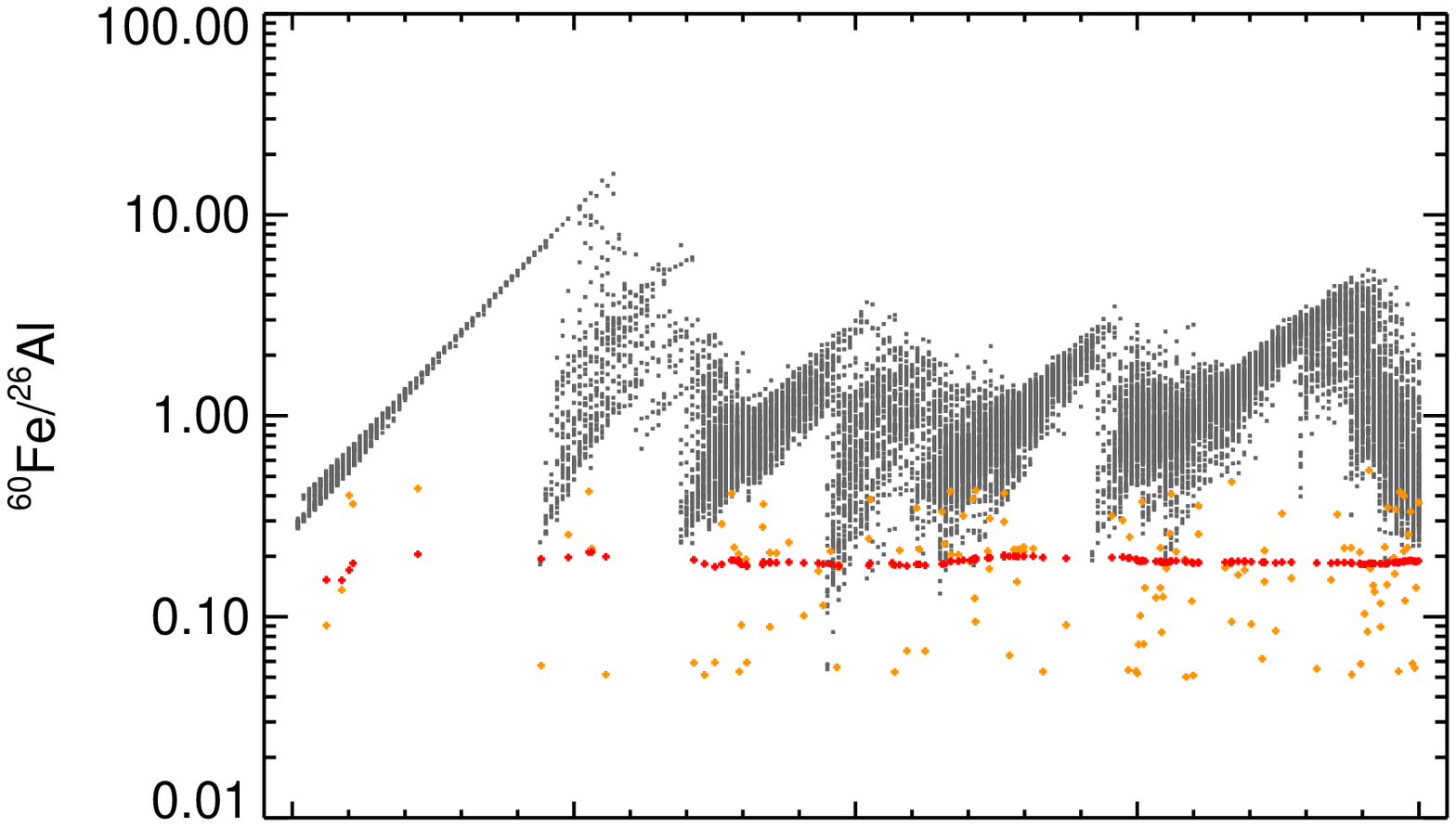}}
  \vspace*{-1.0cm}
  \centerline{\includegraphics[width=\wfig\linewidth]{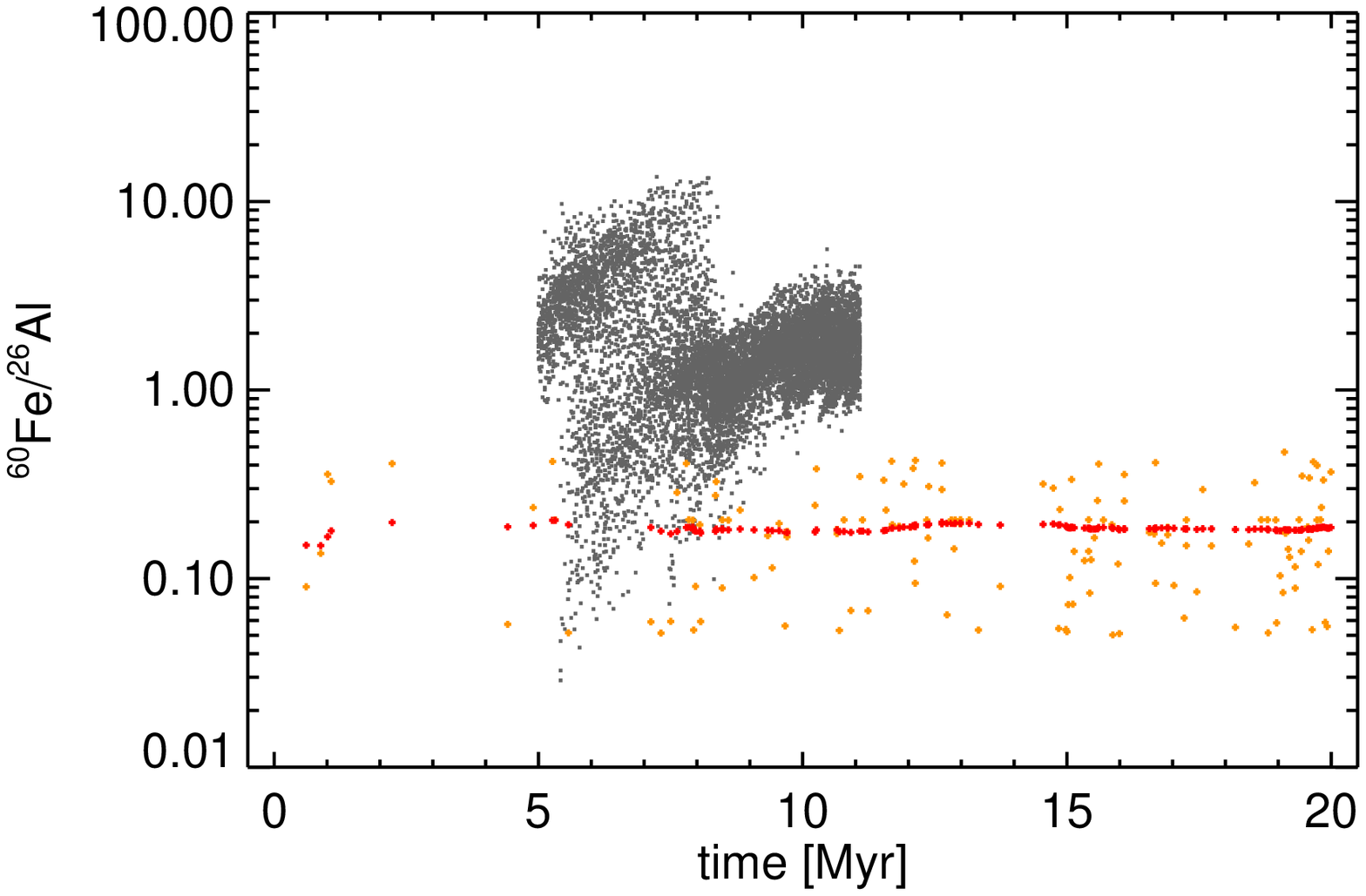}}
  \caption{$^{60}$Fe/$^{26}$Al ratios in the star-forming gas of an evolving GMC for cold star-forming clumps (grey) and individual supernova yields (orange), from low-resolution {\sc stagger} code simulations (top) and from high-resolution {\sc ramses} simulations (bottom), with the ratio of accumulated yields (red).}
  \label{fig:3}
\end{figure}

\Figure{3} shows the ratio of $^{60}$Fe to $^{26}$Al in star-forming gas and individual supernova ejecta, as well as the ratio of accumulated yields at any one time. The latter provides a direct estimate of the ratio of expected $\gamma$-ray line fluxes, since the large majority of ejected radioactive $^{26}$Al and $^{60}$Fe nuclei will have had time to decay. For both the {\sc stagger} and {\sc ramses} simulations, the predicted $^{60}$Fe/$^{26}$Al ratio of total yields is slightly less than 0.2, consistent with the observed $\gamma$-ray line flux ratio of 0.15$\pm$0.06 \citep{2007A&A...469.1005W}.  The instantaneous ratio of average number densities of $^{60}$Fe and $^{26}$Al is larger by the ratio of their half-life, 3.6. Moreover, the average $^{60}$Fe/$^{26}$Al ratio in star-forming gas is expected to be larger than the global average, reflecting the time required for hot supernova ejecta to be converted into dense star-forming gas. Collectively, this implies that the expected average $^{60}$Fe/$^{26}$Al ratio in star forming gas is typically considerably larger than 0.2 (\Figure {3}).

\section{Timescales and mode of admixing supernova ejecta into star-forming gas}

\begin{figure}[htb!]
  \centering
  \includegraphics[width=\wfig\linewidth]{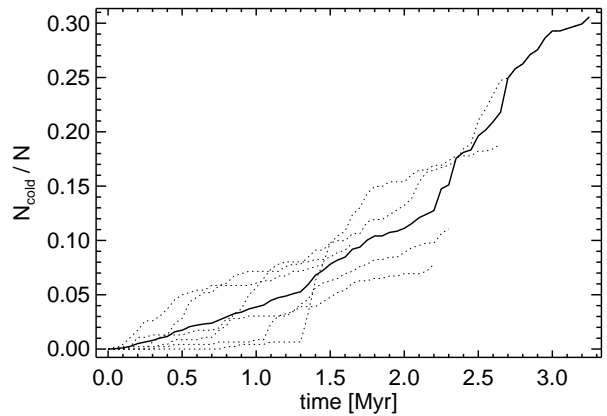}
  \caption{Accumulated histogram of the fraction of supernova ejecta trace particles (N) that has reached a
  cold and dense (N$_{cold}$), potentially star-forming state after a given time. The solid curve is an average over the six individual cases shown by dotted lines. These results were obtained by launching, at the time of supernova explosions, a set of trace particles that travel passively with the ejecta, following them until they were in dense, potentially star-forming gas.}
  \label{fig:4}
\end{figure}

The current perception of transport and mixing of chemical yields in supernova-driven ISM is that this process is rather inefficient, leading to mixing timescales that are longer than the typical lifetime of GMCs \citep{2002ApJ...581.1047D}. However, these simulations are based on the passive seeding of particle tracers to a turbulent medium and, therefore, the tracer particles are unrelated to the supernova events driving the ISM. In contrast, our code allows for linking passive scalars and particle tracers used as proxies for $^{26}$Al and $^{60}$Fe with the kinetic energy of individual supernovae, thereby providing realistic timescales for the transport and admixing of supernova ejecta into star-forming molecular gas. For example, transfer of kinetic energy to the general ISM from slowed-down SN ejecta drives additional compression and conversion to turbulent, cold MC gas, which imparts collapse of some gas fraction within its dynamical timescale. To illustrate the rapid incorporation of supernova-derived $^{26}$Al and $^{60}$Fe into star-forming regions we show in \Figure {4} accumulated fractions of initially hot supernova ejecta that reached cold, dense, and potentially star-forming gas, after a given time. These results indicate that in some cases (of the order 1\%; cf. \Fig{4}), only a few 100,000 years are needed for part of the supernova yields to be incorporated into self-gravitating cores. A significant fraction of supernova ejecta (5--10\%) has been transformed into star-forming gas over a timescale of only 1--2 Myr. 

\begin{figure}[htb!]
  \centering
  \includegraphics[width=\wfigb\linewidth]{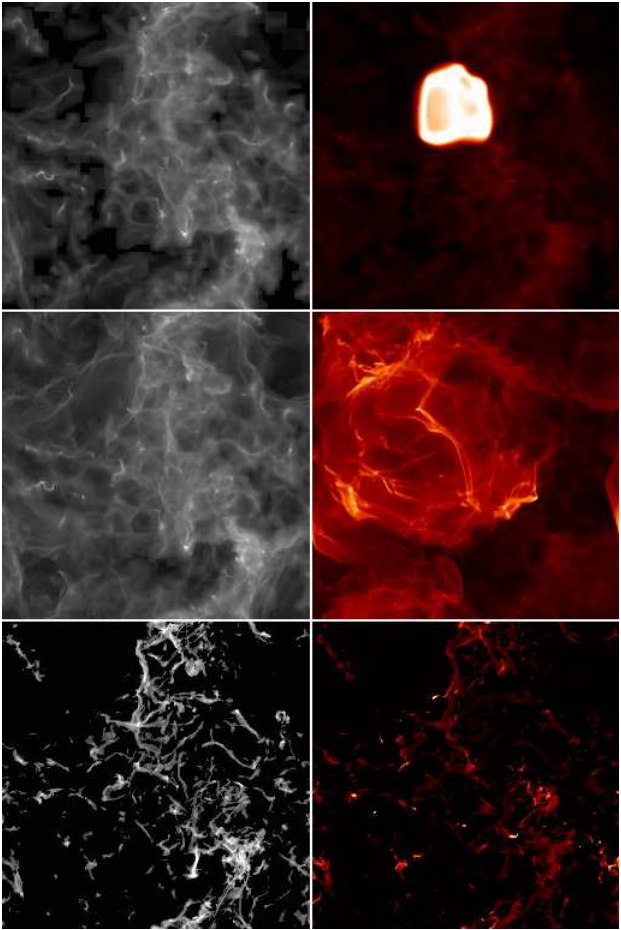}
  \caption{Images of projected average total mass density per unit volume (left) and $2\cdot10^8$ times the $^{26}$Al mass density per unit volume (right), shortly after the explosion of a supernova. The box size is 40 pc. The middle two panels correspond to a time 1 Myr later than the upper two panels, with identical normalization. To enhance visibility the gray scales corresponds to actual values raised to 0.4. The two lowermost panels show the same projections as the middle panels, but with all mass with number density below $3\cdot10^3$ cm$^{-3}$ removed.}
  \label{fig:5}
\end{figure}

To better understand how stellar yields are transported into star-forming regions, we show in Figure \ref{fig:5} examples of projected total and $^{26}$Al number densities during the expansion of a supernova bubble over a time interval of 1 Myr. Star-forming molecular cloud fragments are visible as filamentary structures in the total number density projections, and these do not change dramatically during the 1 Myr period. In contrast, the supernova bubble expands to cover a significant fraction of the gas volume. As illustrated by the right hand side panels, where the upper two panels have the same absolute normalization, supernova ejecta quickly become part of the star-forming cold medium. The lower two panels, which represent the same time interval as the middle panels, show exclusively dense ($N > 3\,10^3$ cm$^{-3}$) and cold (typically less than 300 K) gas, emphasizing that part of the new supernova yields have already been incorporated in that medium. This is because the kinetic energy and supernova yields travel together, becoming progressively admixed into the warm and cold phases of the ISM as the expanding supernova shells exchange momentum with the entrained ISM gas. Moreover, as the kinetic energy from supernovae is likely one of the main contributors to the supersonic and super-Alfv{\'e}nic turbulence that causes protostellar collapse, supernova yields are naturally part of the less dense medium from which pre-stellar cores form. Thus, the presence of freshly-synthesized $^{26}$Al and $^{60}$Fe in star-forming regions does not require external seeding such as hypothesized in models where admixing of short-lived radioisotopes results from the interaction of a supernova shock wave with an un-contaminated pre-stellar core \citep{1977Icar...30..447C,2010ApJ...708.1268B,2012ApJ...756..102P}.

\section{The Solar System's initial $^{\bf60}$Fe/$^{\bf26}$Al ratio}

The level of $^{26}$Al at the time of Solar System formation is well-constrained from the $^{26}$Al-$^{26}$Mg systematics of the oldest Solar System solids, CAIs, which define an initial $^{26}$Al/$^{27}$Al value of $\sim5\cdot10^{-5}$ \citep{2008E&PSL.272..353J,2011ApJ...735L..37L}. This ratio is comparable to the $^{26}$Al/$^{27}$Al values of star-forming gas inferred from our simulations of evolving GMCs, in particular after 10 Myr of evolution, where average $^{26}$Al abundances are similar to that of the nascent Solar System (Fig. 1b). Although there is considerable debate in the literature regarding the initial abundance of $^{60}$Fe at the birth of the Solar System \citep{2006ApJ...639L..87T,2009GeCoA..73.4963K,2012LPI....43.2733T}, a recent study of pristine differentiated meteorites and chondritic components suggest that the initial Solar System $^{60}$Fe/$^{56}$Fe may have been as low as $\sim$1$\cdot$10$^{-8}$ \citep{2012E&PSL.359..248T}. At face value, this implies an initial $^{60}$Fe/$^{26}$Al of $\sim$0.002 \citep[assuming a solar $^{27}$Al/$^{56}$Fe = 0.104,][]{2009ARA&A..47..481A}, which is much lower than the observed $\gamma$-ray flux $^{60}$Fe/$^{26}$Al ratio of $\sim$0.2 as well as that of any star-forming gas sampled from our simulations (Fig. 3). We consider three possibilities for this mismatch: ({\it i}) uncertainties in the $^{60}$Fe supernova yields, ({\it ii}) contamination of the protosolar MC from $^{26}$Al-rich winds of nearby massive or asymptotic giant branch (AGB) stars, and ({\it iii}) the Solar System's initial $^{60}$Fe/$^{56}$Fe value inferred from meteorites sampled so far is not representative of the initial Solar System's bulk $^{60}$Fe abundance.

No plausible modification of the supernovae and stellar winds yields can locally simultaneously give the observed high Solar System level of $^{26}$Al/$^{27}$Al and the very low level of $^{60}$Fe/$^{56}$Fe inferred from meteorites \citep{2012E&PSL.359..248T}. This result is robust and model independent, in that the average ratio of the yields in the simulations agree with the galactic ratio, which is well constrained by $\gamma$-ray observations \citep{2007A&A...469.1005W}.  As per the lower panel in \Fig{3}, local values of $^{60}$Fe/$^{26}$Al below 0.10 are exceedingly rare. Thus, uncertainties in the supernova yields to account for the apparent low Solar System $^{60}$Fe/$^{26}$Al ratio can be ruled out.

Contributions to the gas subsequently forming the Solar System by winds of a massive star or an AGB star could result in the apparent low $^{60}$Fe/$^{26}$Al inferred from some meteoritic studies, given that $^{60}$Fe is not efficiently produced in these stellar environments. However, the space-time window of a star-forming region with gas of this low $^{60}$Fe/$^{26}$Al ratio is small \citep{1994ApJ...421..605K,2010ConPh..51..381W}, requiring an isolated GMC pocket where $^{60}$Fe has decayed away, and is in the neighborhood of a rare, very massive star, just before the end of its brief lifetime, or is in the close neighborhood of a stray AGB star when it emits an $^{26}$Al-rich wind. Although this cannot be ruled out, we estimate from our simulations the probability of such event to be less than 10$^{-6}$.

Alternatively, the low initial Solar System abundance of $^{60}$Fe/$^{56}$Fe inferred from differentiated meteorites \citep{2012E&PSL.359..248T} may not be representative of the average initial abundance of $^{60}$Fe in the Solar System. It has been proposed \citep{2011ApJ...735L..37L} that large-scale heterogeneity in the initial abundance of $^{26}$Al may have existed throughout the inner Solar System, with possibly up to 80 \% reduction relative to the canonical initial $^{26}$Al/$^{27}$Al ratio defined by CAIs. The $^{26}$Al heterogeneity amongst inner Solar System objects has been ascribed to thermal processing of molecular cloud material \citep{2009Sci...324..374T,2013ApJ...763L..40P}, which resulted in preferential loss by sublimation of a more volatile $^{26}$Al-rich carrier, producing residual isotopic heterogeneity. Thus, CAIs represent samples of the complementary gaseous reservoir enriched in $^{26}$Al by thermal processing, which resulted in the widespread $^{26}$Al depletions observed among the inner Solar System bodies. If the $^{60}$Fe carrier phase was significantly more volatile than the $^{26}$Al carrier, then thermal processing would have effectively removed most of the $^{60}$Fe from the inner Solar System solids, leading to a significant enhancement of the $^{60}$Fe concentration of the gaseous phase at that time. If this interpretation is correct, we predict that CAIs, provided they are samples of a complementary gas reservoir, will define internal isochron relationships consistent with initial $^{60}$Fe/$^{56}$Fe ratios of the order 10$^{-6}$.

The close agreement of the Sun's abundance pattern with that of solar-twins in the M67 open cluster \citep{2011A&A...528A..85O} suggest that our Solar System may have formed in a GMC giving rise to a similar high-mass cluster. The high $^{26}$Al and $^{60}$Fe concentrations in star-forming gas indicated by our simulations of a comparable environment suggest that the presence of short-lived radionuclides in the early Solar System does not require an unusual formation scenario. Direct measurement of a $^{60}$Fe/$^{26}$Al ratio in CAIs that is comparable to that deduced from our GMC simulations would give much credence to this proposal.


\vspace{12pt}
Financed by the Danish National Research Foundation (grant number DNRF97). Computing resources provided by the Danish Center for Scientific Computing and by the NASA High-End Computing Capability at Ames.


\begin{thebibliography}{41}
\expandafter\ifx\csname natexlab\endcsname\relax\def\natexlab#1{#1}\fi

\bibitem[{{Asplund} {et~al.}(2009){Asplund}, {Grevesse}, {Sauval}, \&
  {Scott}}]{2009ARA&A..47..481A}
{Asplund}, M., {Grevesse}, N., {Sauval}, A.~J., \& {Scott}, P. 2009, \araa, 47,
  481

\bibitem[{{Boss} {et~al.}(2010){Boss}, {Keiser}, {Ipatov}, {Myhill}, \&
  {Vanhala}}]{2010ApJ...708.1268B}
{Boss}, A.~P., {Keiser}, S.~A., {Ipatov}, S.~I., {Myhill}, E.~A., \& {Vanhala},
  H.~A.~T. 2010, \apj, 708, 1268

\bibitem[{{Cameron} \& {Truran}(1977)}]{1977Icar...30..447C}
{Cameron}, A.~G.~W., \& {Truran}, J.~W. 1977, \icarus, 30, 447

\bibitem[{{Chabrier}(2005)}]{2005ASSL..327...41C}
{Chabrier}, G. 2005, in Astrophysics and Space Science Library, Vol. 327, The
  Initial Mass Function 50 Years Later, ed. E.~{Corbelli}, F.~{Palla}, \&
  H.~{Zinnecker}, 41

\bibitem[{{de Avillez} \& {Mac Low}(2002)}]{2002ApJ...581.1047D}
{de Avillez}, M.~A., \& {Mac Low}, M.-M. 2002, \apj, 581, 1047

\bibitem[{{Diehl} {et~al.}(2006){Diehl}, {Halloin}, {Kretschmer}, {Lichti},
  {Sch{\"o}nfelder}, {Strong}, {von Kienlin}, {Wang}, {Jean}, {Kn{\"o}dlseder},
  {Roques}, {Weidenspointner}, {Schanne}, {Hartmann}, {Winkler}, \&
  {Wunderer}}]{2006Natur.439...45D}
{Diehl}, R., {Halloin}, H., {Kretschmer}, K., {Lichti}, G.~G.,
  {Sch{\"o}nfelder}, V., {Strong}, A.~W., {von Kienlin}, A., {Wang}, W.,
  {Jean}, P., {Kn{\"o}dlseder}, J., {Roques}, J.-P., {Weidenspointner}, G.,
  {Schanne}, S., {Hartmann}, D.~H., {Winkler}, C., \& {Wunderer}, C. 2006,
  \nat, 439, 45

\bibitem[{{Elmegreen}(2000)}]{2000ApJ...530..277E}
{Elmegreen}, B.~G. 2000, \apj, 530, 277

\bibitem[{{Elmegreen}(2013)}]{2013IAUS..292...35E}
{Elmegreen}, B.~G. 2013, in IAU Symposium, Vol. 292, IAU Symposium, 35--38

\bibitem[{{Franco} \& {Cox}(1986)}]{1986PASP...98.1076F}
{Franco}, J., \& {Cox}, D.~P. 1986, \pasp, 98, 1076

\bibitem[{{Fromang} {et~al.}(2005){Fromang}, {Hennebelle}, \&
  {Teyssier}}]{2005sf2a.conf..743F}
{Fromang}, S., {Hennebelle}, P., \& {Teyssier}, R. 2005, in SF2A-2005: Semaine
  de l'Astrophysique Francaise, ed. F.~{Casoli}, T.~{Contini}, J.~M. {Hameury},
  \& L.~{Pagani}, 743

\bibitem[{{Gnedin} \& {Hollon}(2012)}]{2012ApJS..202...13G}
{Gnedin}, N.~Y., \& {Hollon}, N. 2012, \apjs, 202, 13

\bibitem[{{Goswami}(2004)}]{2004NewAR..48..125G}
{Goswami}, J.~N. 2004, New Astronomy, 48, 125

\bibitem[{{Huss} {et~al.}(2009){Huss}, {Meyer}, {Srinivasan}, {Goswami}, \&
  {Sahijpal}}]{2009GeCoA..73.4922H}
{Huss}, G.~R., {Meyer}, B.~S., {Srinivasan}, G., {Goswami}, J.~N., \&
  {Sahijpal}, S. 2009, \gca, 73, 4922

\bibitem[{{Jacobsen} {et~al.}(2008){Jacobsen}, {Yin}, {Moynier}, {Amelin},
  {Krot}, {Nagashima}, {Hutcheon}, \& {Palme}}]{2008E&PSL.272..353J}
{Jacobsen}, B., {Yin}, Q.-Z., {Moynier}, F., {Amelin}, Y., {Krot}, A.~N.,
  {Nagashima}, K., {Hutcheon}, I.~D., \& {Palme}, H. 2008, Earth and Planetary
  Science Letters, 272, 353

\bibitem[{{Kastner} \& {Myers}(1994)}]{1994ApJ...421..605K}
{Kastner}, J.~H., \& {Myers}, P.~C. 1994, \apj, 421, 605

\bibitem[{{Kennicutt} \& {Evans}(2012)}]{2012ARA&A..50..531K}
{Kennicutt}, R.~C., \& {Evans}, N.~J. 2012, \araa, 50, 531

\bibitem[{{Kritsuk} {et~al.}(2011){Kritsuk}, {Nordlund}, {Collins}, {Padoan},
  {Norman}, {Abel}, {Banerjee}, {Federrath}, {Flock}, {Lee}, {Li},
  {M{\"u}ller}, {Teyssier}, {Ustyugov}, {Vogel}, \& {Xu}}]{2011ApJ...737...13K}
{Kritsuk}, A.~G., {Nordlund}, {\AA}., {Collins}, D., {Padoan}, P., {Norman},
  M.~L., {Abel}, T., {Banerjee}, R., {Federrath}, C., {Flock}, M., {Lee}, D.,
  {Li}, P.~S., {M{\"u}ller}, W.-C., {Teyssier}, R., {Ustyugov}, S.~D., {Vogel},
  C., \& {Xu}, H. 2011, \apj, 737, 13

\bibitem[{{Kritsuk} \& {Norman}(2011)}]{2011arXiv1111.2827K}
{Kritsuk}, A.~G., \& {Norman}, M.~L. 2011, ArXiv e-prints

\bibitem[{{Krot} {et~al.}(2009){Krot}, {Amelin}, {Bland}, {Ciesla}, {Connelly},
  {Davis}, {Huss}, {Hutcheon}, {Makide}, {Nagashima}, {Nyquist}, {Russell},
  {Scott}, {Thrane}, {Yurimoto}, \& {Yin}}]{2009GeCoA..73.4963K}
{Krot}, A.~N., {Amelin}, Y., {Bland}, P., {Ciesla}, F.~J., {Connelly}, J.,
  {Davis}, A.~M., {Huss}, G.~R., {Hutcheon}, I.~D., {Makide}, K., {Nagashima},
  K., {Nyquist}, L.~E., {Russell}, S.~S., {Scott}, E.~R.~D., {Thrane}, K.,
  {Yurimoto}, H., \& {Yin}, Q.-Z. 2009, \gca, 73, 4963

\bibitem[{{Krumholz} \& {McKee}(2005)}]{2005ApJ...630..250K}
{Krumholz}, M.~R., \& {McKee}, C.~F. 2005, \apj, 630, 250

\bibitem[{{Larsen} {et~al.}(2011){Larsen}, {Trinquier}, {Paton}, {Schiller},
  {Wielandt}, {Ivanova}, {Connelly}, {Nordlund}, {Krot}, \&
  {Bizzarro}}]{2011ApJ...735L..37L}
{Larsen}, K.~K., {Trinquier}, A., {Paton}, C., {Schiller}, M., {Wielandt}, D.,
  {Ivanova}, M.~A., {Connelly}, J.~N., {Nordlund}, {\AA}., {Krot}, A.~N., \&
  {Bizzarro}, M. 2011, \apjl, 735, L37

\bibitem[{{Larson}(1981)}]{1981MNRAS.194..809L}
{Larson}, R.~B. 1981, \mnras, 194, 809

\bibitem[{{Limongi} \& {Chieffi}(2006)}]{2006ApJ...647..483L}
{Limongi}, M., \& {Chieffi}, A. 2006, \apj, 647, 483

\bibitem[{{Meynet} {et~al.}(1994){Meynet}, {Maeder}, {Schaller}, {Schaerer}, \&
  {Charbonnel}}]{1994A&AS..103...97M}
{Meynet}, G., {Maeder}, A., {Schaller}, G., {Schaerer}, D., \& {Charbonnel}, C.
  1994, \aaps, 103, 97

\bibitem[{{Murray}(2011)}]{2011ApJ...729..133M}
{Murray}, N. 2011, \apj, 729, 133

\bibitem[{{Norris} {et~al.}(1983){Norris}, {Gancarz}, {Rokop}, \&
  {Thomas}}]{1983JGR....88..331N}
{Norris}, T.~L., {Gancarz}, A.~J., {Rokop}, D.~J., \& {Thomas}, K.~W. 1983,
  \jgr, 88, 331

\bibitem[{{{\"O}nehag} {et~al.}(2011){{\"O}nehag}, {Korn}, {Gustafsson},
  {Stempels}, \& {Vandenberg}}]{2011A&A...528A..85O}
{{\"O}nehag}, A., {Korn}, A., {Gustafsson}, B., {Stempels}, E., \&
  {Vandenberg}, D.~A. 2011, \aap, 528, A85

\bibitem[{{Osterbrock}(1989)}]{1989agna.book.....O}
{Osterbrock}, D.~E. 1989, {Astrophysics of gaseous nebulae and active galactic
  nuclei}

\bibitem[{{Padoan} {et~al.}(2012){Padoan}, {Haugb{\o}lle}, \&
  {Nordlund}}]{2012ApJ...759L..27P}
{Padoan}, P., {Haugb{\o}lle}, T., \& {Nordlund}, {\AA}. 2012, \apjl, 759, L27

\bibitem[{{Padoan} \& {Nordlund}(2002)}]{2002ApJ...576..870P}
{Padoan}, P., \& {Nordlund}, {\AA}. 2002, \apj, 576, 870

\bibitem[{{Padoan} \& {Nordlund}(2011)}]{2011ApJ...730...40P}
---. 2011, \apj, 730, 40

\bibitem[{{Pan} {et~al.}(2012){Pan}, {Desch}, {Scannapieco}, \&
  {Timmes}}]{2012ApJ...756..102P}
{Pan}, L., {Desch}, S.~J., {Scannapieco}, E., \& {Timmes}, F.~X. 2012, \apj,
  756, 102

\bibitem[{{Paton} {et~al.}(2013){Paton}, {Schiller}, \&
  {Bizzarro}}]{2013ApJ...763L..40P}
{Paton}, C., {Schiller}, M., \& {Bizzarro}, M. 2013, \apjl, 763, L40

\bibitem[{{Rugel} {et~al.}(2009){Rugel}, {Faestermann}, {Knie}, {Korschinek},
  {Poutivtsev}, {Schumann}, {Kivel}, {G{\"u}nther-Leopold}, {Weinreich}, \&
  {Wohlmuther}}]{2009PhRvL.103g2502R}
{Rugel}, G., {Faestermann}, T., {Knie}, K., {Korschinek}, G., {Poutivtsev}, M.,
  {Schumann}, D., {Kivel}, N., {G{\"u}nther-Leopold}, I., {Weinreich}, R., \&
  {Wohlmuther}, M. 2009, Physical Review Letters, 103, 072502

\bibitem[{{Tachibana} {et~al.}(2006){Tachibana}, {Huss}, {Kita}, {Shimoda}, \&
  {Morishita}}]{2006ApJ...639L..87T}
{Tachibana}, S., {Huss}, G.~R., {Kita}, N.~T., {Shimoda}, G., \& {Morishita},
  Y. 2006, \apjl, 639, L87

\bibitem[{{Tang} \& {Dauphas}(2012)}]{2012E&PSL.359..248T}
{Tang}, H., \& {Dauphas}, N. 2012, Earth and Planetary Science Letters, 359,
  248

\bibitem[{{Telus} {et~al.}(2012){Telus}, {Huss}, {Nagashima}, {Ogliore}, \&
  {Tachibana}}]{2012LPI....43.2733T}
{Telus}, M., {Huss}, G.~R., {Nagashima}, K., {Ogliore}, R.~C., \& {Tachibana},
  S. 2012, in Lunar and Planetary Inst. Technical Report, Vol.~43, Lunar and
  Planetary Institute Science Conference Abstracts, 2733

\bibitem[{{Teyssier}(2002)}]{2002A&A...385..337T}
{Teyssier}, R. 2002, \aap, 385, 337

\bibitem[{{Trinquier} {et~al.}(2009){Trinquier}, {Elliott}, {Ulfbeck}, {Coath},
  {Krot}, \& {Bizzarro}}]{2009Sci...324..374T}
{Trinquier}, A., {Elliott}, T., {Ulfbeck}, D., {Coath}, C., {Krot}, A.~N., \&
  {Bizzarro}, M. 2009, Science, 324, 374

\bibitem[{{Wang} {et~al.}(2007){Wang}, {Harris}, {Diehl}, {Halloin}, {Cordier},
  {Strong}, {Kretschmer}, {Kn{\"o}dlseder}, {Jean}, {Lichti}, {Roques},
  {Schanne}, {von Kienlin}, {Weidenspointner}, \&
  {Wunderer}}]{2007A&A...469.1005W}
{Wang}, W., {Harris}, M.~J., {Diehl}, R., {Halloin}, H., {Cordier}, B.,
  {Strong}, A.~W., {Kretschmer}, K., {Kn{\"o}dlseder}, J., {Jean}, P.,
  {Lichti}, G.~G., {Roques}, J.~P., {Schanne}, S., {von Kienlin}, A.,
  {Weidenspointner}, G., \& {Wunderer}, C. 2007, \aap, 469, 1005


\bibitem[{{Wielandt} {et~al.}(2012){Wielandt}, {Nagashima}, {Krot}, {Huss}, 
  {Ivanova} \& {Bizzarro}}]{2012ApJ...748L..25W}
{Wielandt}, D., {Nagashima}, K., {Krot}, A.~N., {Huss}, G.~R., {Ivanova}, M.~A.,
   \& {Bizzarro}, M. 2012, \apjl, 748, L25


\bibitem[{{Williams}(2010)}]{2010ConPh..51..381W}
{Williams}, J. 2010, Contemporary Physics, 51, 381






\end{thebibliography}

\end{document}